\documentclass[aps,nofootinbib,superscriptaddress,twocolumn,floatfix]{revtex4}
\usepackage{amsmath,amsfonts,booktabs,csvsimple,dcolumn,doi,float,graphicx,makecell,physics,url,wrapfig}
\usepackage[version=4]{mhchem}

\allowdisplaybreaks

\usepackage[separate-uncertainty=true,separate-uncertainty-units=single,input-digits=0123456789\pi]{siunitx}
\AtBeginDocument{\RenewCommandCopy\qty\SI}
\DeclareSIUnit\angstrom{\text{Å}}

\newcommand{\kummer}[3]{M\left(#1,#2,#3\right)}

\usepackage{ulem}   

\usepackage{xcolor}
\newcommand{\FM}[1]{\textcolor{black}{#1}}

\begin{document}

\title{Analytical Solution to the Kronig-Penney Model with Harmonic Oscillator Wells: Insights to Tight-Binding}
\author{Christopher Moore}
\affiliation{Department of Physics, University of Alberta, Edmonton, AB, Canada T6G~2E1}
\author{Frank Marsiglio}
 \affiliation{Department of Physics, University of Alberta, Edmonton, AB, Canada T6G~2E1}
 \affiliation{Theoretical Physics Institute \& Quantum Horizons Alberta, University of Alberta, Edmonton, Alberta T6G 2E1, Canada}


\begin{abstract}
The celebrated Kronig-Penney model traditionally has been formulated with square well potentials representing atomic centres. Here, we use a slightly more realistic potential, the truncated harmonic oscillator, in lieu of square well potentials, and solve the model analytically. We derive the energy dispersion and wave functions for this model. This configuration has some important similarities and differences compared to the usual model. In particular, we write the governing equation in a form suggestive of the tight-binding approximation, as can be done for the usual model. In this way, it is straightforward to derive an expression for the tunneling amplitude used in tight-binding in terms of the harmonic oscillator potential parameters. 
\end{abstract}

\maketitle

\section{Introduction}

The theory of the electronic structure of solids arguably started with Felix Bloch in 1929~\cite{Bloch} (see also \cite{bloch1980} for a perspective on his earlier work). Not long thereafter, this work was followed by seminal papers in 1930 by Brillouin \cite{brillouin1930} and in 1931 by Kronig and Penney~\cite{Kronig-Penney}.
The Kronig-Penney model consists of a one-dimensional periodic array of potential square wells meant to represent the atomic potentials in a lattice. The solution for this model consists of energy bands separated by gaps in the spectrum \cite{Kronig-Penney}.
This paper, followed by the Wigner-Seitz papers on metallic Sodium \cite{wigner1933,wigner1934}, served not only to establish a methodology for distinguishing metals from band insulators, but for also calculating specific details of electronic band structure. Of course, the latter goal requires the use of more realistic Coulombic potentials in three-dimensional lattices \cite{martin2020}.

In parallel to the development of a more realistic description of the behaviour of electrons in a solid, a number of papers have focussed on more qualitative aspects of the original Kronig-Penney model 
\cite{johnston92,Kronig-Penney_Matrix_Mechanics,pavelich16,johnston2019,johnston2020,Kronig-Penney_Tight_Binding,forcade21}. These  papers generally use numerical methods to solve for the dispersion relations and wavefunctions in a number of variations on the original Kronig-Penney model.

\begin{figure}
    \centering
    \includegraphics[width=\columnwidth]{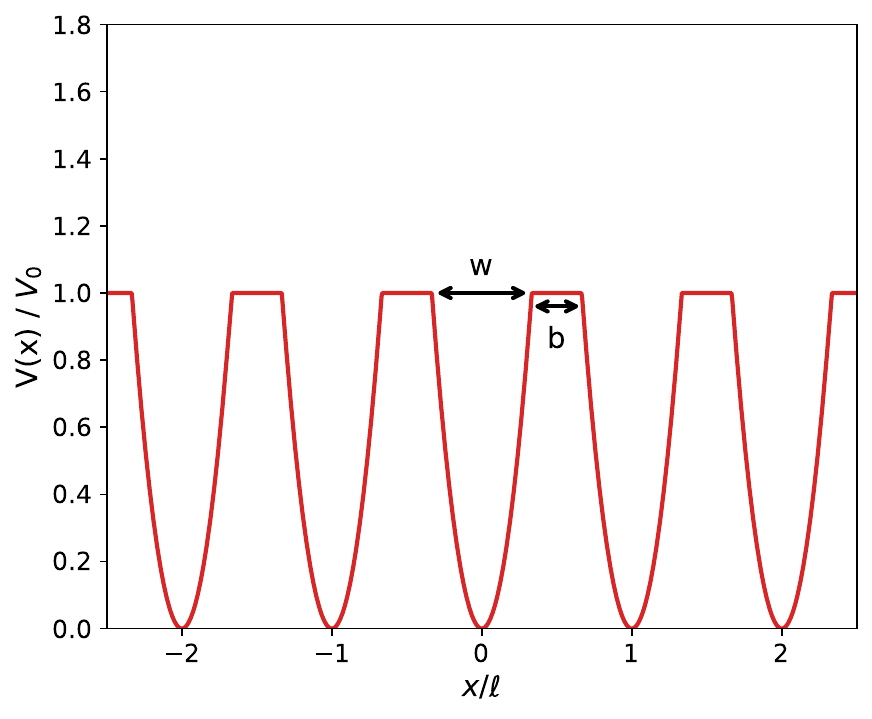}
    \caption{Diagram of a one-dimensional periodic potential with atomic sites surrounded by harmonic oscillator potential wells, and with horizontal plateaus representing the barriers between atoms. In this case we use a unit cell length $\ell = 1$, with well-width $w = 2\ell/3$, and barrier widths $b/2 = \ell/6$ on either side of each well.}
    \label{fig:Oscillator Plateau Potential}
\end{figure}

In this paper we will instead investigate analytic solutions for a Kronig-Penney-like model using harmonic oscillator potentials in place of square wells, separated by flat barrier regions, as illistrated in Fig.~\ref{fig:Oscillator Plateau Potential}. \FM{This is a slightly more realistic model for atoms, as electrons prefer to be closer to the centre of the atomic potential.} An analytical solution is still possible in this case, with the use of non-elementary functions. The analytical structure of the original Kronig-Penney model solution is retained and, in this way, one can note commonalities and differences with the original result. In particular, the tight-binding limit is readily derived, and illustrates features that are similar to the original. This work complements existing work done using numerical matrix mechanics but, because it is analytical, no limitations arise from Hilbert Space truncation and other restrictions imposed by the requirement to diagonalize large matrices \cite{Kronig-Penney_Matrix_Mechanics}.

The outline is as follows. In Sec.~\ref{sec:bcs} we begin with a general solution of the truncated harmonic oscillator potential. Most of the details are relegated to Appendix~\ref{sec:Infinite Well Derivation}. This is followed by a derivation of the Kronig-Penney equations for harmonic oscillator potentials in Sec.~\ref{sec:bloch bc}, and these are written in a form suggestive of a tight-binding model. Sec.~\ref{sec:tight} describes the derivation of a tight-binding dispersion relation, with an expression for the tunneling matrix element in terms of the parameters of the underlying potential. We end with conclusions in Sec.~\ref{conc},
and include a second Appendix \ref{sec:Numerical Solutions} to describe the numerical solutions, for completeness.




\section{The importance of boundary conditions}
\label{sec:bcs}
Because solutions to truncated harmonic oscillator potentials are less familiar to readers, it is important to proceed in steps. We start with the isolated harmonic potential, where
\begin{equation}\label{eq:Oscillator Plateau Potential}
    V_\mathrm{HO}(x) = V_0\begin{cases}
        1 & \text{if } \  -\infty \ \ < x \leq -w/2 \\
        (\frac{x}{w/2})^2 & \text{if }\ -w/2 \leq x \leq \ \ w/2 \ \\
        1 & \text{if }\ \ \ \ \ w/2 \leq x < \ \ \infty.
    \end{cases}
\end{equation}
Note that 
\begin{equation}
    V_0 = \frac{1}{2} m \omega^2 \left(\frac{w}{2}\right)^2 = \left( \frac{w}{2x_0} \right)^2 \frac{\hbar \omega}{2} \equiv v_0 \frac{\hbar \omega}{2},
    \label{harm_pot}
\end{equation}
where in the last two equalities we have written the potential in natural units of $\hbar \omega/2$ and we have defined the dimensionless potential $v_0$. Here, $x_0 \equiv \sqrt{\hbar/(m\omega)}$ is the length scale associated with the usual harmonic oscillator potential. This potential corresponds to the single central well in Fig.~\ref{fig:Oscillator Plateau Potential}, but with the horizontal barriers extending to $\pm \infty$. 

The details in the derivation are provided in Appendix~\ref{sec:Infinite Well Derivation}. The resulting wave function, for states with $E<V_0$, is given by
\begin{widetext}
\begin{equation}\label{eq:Oscillator Plateau Wavefunction}
    \psi(x) = \begin{cases}
        A \exp(\kappa \left(x + \frac{w}{2}\right)) & \text{if } \ \ -\infty < x \leq -w/2 \\
        \left(C \kummer{\frac{1 - \epsilon}{4}}{\frac{1}{2}}{[x/x_0]^2} + D \frac{x}{x_0} \kummer{\frac{3 - \epsilon}{4}}{\frac{3}{2}}{[x/x_0]^2}\right) e^{-[x/x_0]^2/2} & \text{if } -w/2 \leq x \leq +w/2 \\
        F \exp(\kappa \left(\frac{w}{2} - x\right)) & \text{if } \  +w/2 \leq x < +\infty \\
    \end{cases} \\
\end{equation}
\end{widetext}
where \(\kummer{a}{b}{z}\) is the Kummer function~\cite{Handbook_Math_Funcs}, and \mbox{$\epsilon \equiv E/(\hbar \omega/2)$} and $\kappa \equiv \sqrt{\frac{2m}{\hbar^2}(V_0 - E)}$. The latter expression can also be written as $\kappa x_0 = \sqrt{v_0 - \epsilon}$.
The two terms in the central portion of this solution correspond to the two independent solutions of the confluent hypergeometric equation. Their properties are well documented \cite{Handbook_Math_Funcs,NIST2020}, and they play the role of the $\approx \sin{(x/w)}$ and $\approx \cos{(x/w)}$ solutions in the regular Kronig-Penney model. In fact, it is explicitly implied in Eq.~\ref{eq:Oscillator Plateau Wavefunction} that the first term is even and second term is odd in $x$. Since the potential defined in Eq.~\ref{harm_pot} is even in $x$, the independent solutions can be chosen to be the even and odd solutions. These correspond to $F=A$, $D=0$ (even) and $F=-A$, $C=0$ (odd), respectively.

As shown in Appendix~\ref{sec:Infinite Well Derivation}, matching these wave functions and their derivatives at $x = \pm w/2$ results in the equations
\begin{equation}
f_{\rm even}(\epsilon) \equiv 1 - \frac{1 - (1-\epsilon)m_{53}(\epsilon)}{\sqrt{1 - {\epsilon}/{v_0}}} = 0 \ \ \ \ \ \ \ \ \ \ \ \ \ \ \ \ \ {\rm (even) \ \ \ }
\label{feven}
\end{equation}
and
\begin{equation}
f_{\rm odd}(\epsilon) \equiv 1 - v_0 + v_0\sqrt{1 - \frac{\epsilon}{v_0}} + v_0(1 - \frac{\epsilon}{3})m_{75}(\epsilon) = 0 \ \ \ {\rm (odd),}
\label{fodd}
\end{equation}
where $m_{ij}(\epsilon) \equiv M_{ij}(\epsilon)/M_{i-4,j-2}(\epsilon)$ and $M_{ij}(\epsilon) \equiv M(\frac{i-\epsilon}{4},\frac{j}{2},v_0)$.
These equations were solved by stepping through $\epsilon$ until a zero crossing was observed, and then refining the location of the zero to a desired error level. Note that we found it best to use equations with no $M_{ij}$ in the denominators, as this avoids complications from the poles of the Kummer functions.
We have defined the functions $f_{\rm even}(\epsilon)$ and $f_{\rm odd}(\epsilon)$ for future use. The accuracy of these solutions was confirmed by comparison with the usual harmonic oscillator potential in the limit as $V_0 \rightarrow \infty$, but also through comparisons with a numerical method further discussed in Appendix~\ref{sec:Numerical Solutions}.

Two more related but preliminary problems to the one just discussed are as follows. First one imagines enclosing this potential in an infinite square well of width $\ell$, with the harmonic oscillator potential situated in the middle. Then the solution is required to approach zero at $x = \pm \ell/2$. Since the solution just presented decays exponentially away from $x = \pm w/2$, the presence of the wall will have little effect on the solutions with $E\ll V_0$. 

A second problem is to use periodic boundary conditions instead of Dirichlet boundary conditions, and the outcome is the same. When $E\ll V_0$, the results are essentially identical as before, although the details of both the analytical and numerical treatments are different, as outlined in Appendix~\ref{sec:Infinite Well Derivation}. Stepping through these two cases, however, allows one to proceed straightforwardly with the case applicable to the infinite array depicted in Fig.~\ref{fig:Oscillator Plateau Potential}, which requires the Bloch boundary condition, as is shown in the next section.

\section{Bloch Boundary Conditions}
\label{sec:bloch bc}

In the case of the infinite array, Bloch's Theorem allows one to focus only on one unit cell (say, the central one in Fig.~\ref{fig:Oscillator Plateau Potential}), but with the boundary condition
\begin{equation}
    \psi(x+\ell) = e^{ik\ell} \psi(x)
    \label{eq:bloch_condition}
\end{equation}
where $k$ is the Bloch wave vector. In this case, the wave function in the three regions is given for $E<V_0$ by
\begin{widetext}
\begin{align}
    \label{eq:Oscillator Plateau Full Wavefunction}
    \psi(x) &= \begin{cases}
        A \sinh(\kappa \left(x + \frac{\ell}{2}\right)) + B \cosh(\kappa \left(x + \frac{\ell}{2}\right)) & \text{if } \ -\ell/2 \leq x \leq -w/2 \\
        \left(C \kummer{\frac{1 - \epsilon}{4}}{\frac{1}{2}}{[x/x_0]^2} + D \frac{x}{x_0} \kummer{\frac{3 - \epsilon}{4}}{\frac{3}{2}}{[x/x_0]^2}\right) e^{-[x/x_0]^2/2} & \text{if } -w/2 \leq x \leq w/2 \\
        F \sinh(\kappa \left(\frac{\ell}{2} - x\right)) + G \cosh(\kappa \left(\frac{\ell}{2} - x\right)) & \text{if } \ \ \ \ w/2 \leq x \leq \ell/2. \\
    \end{cases}
\end{align}
\end{widetext}
\noindent A similar expression applies for $E > V_0$. We then apply the Bloch conditions to give $F=-e^{ik\ell}A$ and $G = e^{ik\ell}B$. Then we use the boundary conditions at $x = \pm w/2$ to further relate the coefficients to one another, with the result that a non-trivial solution for the eigenenergy \mbox{$E<V_0$} (a similar one exists for $E>V_0$) is obtained through the transcendental equation
\begin{widetext}
\begin{equation}
\cos(kl) = \frac{X_{75} - X_{53}}{X_{75} + X_{53}}\cosh(\kappa b) + \frac{\sqrt{v_0(v_0 - \epsilon)}}{X_{75} + X_{53}}\left(1 - \frac{X_{53}X_{75}}{v_0(v_0 - \epsilon)}\right)\sinh(\kappa b)
    \label{kp1}
\end{equation}

\noindent where
\begin{align}
    X_{53} & \equiv v_0\left[1 - (1 - \epsilon)m_{53}(\epsilon)\right] \\
    X_{75} & \equiv 1 - v_0 + v_0\left(1 - \frac{\epsilon}{3}\right)m_{75}(\epsilon).
\end{align}
\end{widetext}
Details of this derivation are included in Appendix~\ref{sec:Infinite Well Derivation}. Equation~(\ref{kp1}) is written in this way to strike some resemblance to the way the usual Kronig-Penney model equation is written\FM{,
\begin{equation}
\cos{k\ell} = \cos{q_1 w} \cosh{\kappa b} + \frac{\kappa^2 - q_1^2}{2q_1 \kappa} \sin{q_1w} \sinh{\kappa b},
\label{usual_kp}
\end{equation}
where square wells of width $w$ take the place of the harmonic oscillator potentials in Fig.~\ref{fig:Oscillator Plateau Potential} and $q_1 \equiv \sqrt{2mE/\hbar^2}$.}
(see Eq.~(5) in Ref.~(\cite{forcade21}), for example). In particular, while \FM{either Eq.~(\ref{kp1}) or Eq.~(\ref{usual_kp}) are clearly difficult non-linear equations 
} to solve for solutions $\epsilon(k)$, it is trivial to find $k(\epsilon)$, and simply requires one to evaluate the inverse cosine of the right-hand side. Evaluating $k(\epsilon)$ is equally good for constructing the energy diagram and the resulting wave functions.

A more enlightening way to write Eq.~(\ref{kp1}) is
\begin{equation}
\cos(kl) = \frac{1}{X_{75} + X_{53}}\left(f_{\rm even}(\epsilon) f_{\rm odd}(\epsilon) \sinh(\kappa b) + e^{-\kappa b}\right),
    \label{kp2}
\end{equation}
where the functions $f_{\rm even}$ and $f_{\rm odd}$ are the same functions that appeared in the solution for the single truncated harmonic oscillator well, Eqs.~(\ref{feven},\ref{fodd}). Since
\mbox{$\kappa b=\frac{2}{\ell -b} bv_0 \sqrt{1 - \epsilon/v_0}$}, 
then when the potential barrier is large compared with the energy, or wide with respect to the width of the well, the prefactor $sinh$ function will be exponentially large. All other terms are of order unity or exponentially small, which requires that one of the $f$-functions will be close to zero, which is the condition satisfied by the single well problem. We will exploit this observation in more detail in Section~\ref{sec:tight} below.

\FM{We should note that Eq.~(\ref{usual_kp}) can also be written in an equally enlightening way for the square well potentials \cite{Kronig-Penney_Matrix_Mechanics}, as
\begin{equation}
\label{usual_kp2}
    \cos(kl) = f_{\rm even}^{\rm sq}(\epsilon) f_{\rm odd}^{\rm sq}(\epsilon) \sinh(\kappa b) + e^{-\kappa b}\cos{(q_1 w)},
\end{equation}
where 
\begin{equation}
f_{\rm even}^{\rm sq}(\epsilon) \equiv \cos{(\frac{q_1 w}{2})} - \frac{q_1}{\kappa}\sin{(\frac{q_1 w}{2})}
\label{feven_sq}
\end{equation}
and
\begin{equation}
f_{\rm odd}^{\rm sq}(\epsilon) \equiv \cos{(\frac{q_1 w}{2})} + \frac{\kappa}{q_1} \sin{(\frac{q_1 w}{2})}.
\label{fodd_sq}
\end{equation}
Note that $f_{\rm even}^{\rm sq}(\epsilon_{0\ {\rm even}}) = 0$ and $f_{\rm odd}^{\rm sq}(\epsilon_{0\ {\rm odd}}) = 0$ give the solutions to the single isolated finite well in the even and odd case, respectively.
}

\begin{figure}
    \centering
    \includegraphics[width=\columnwidth]{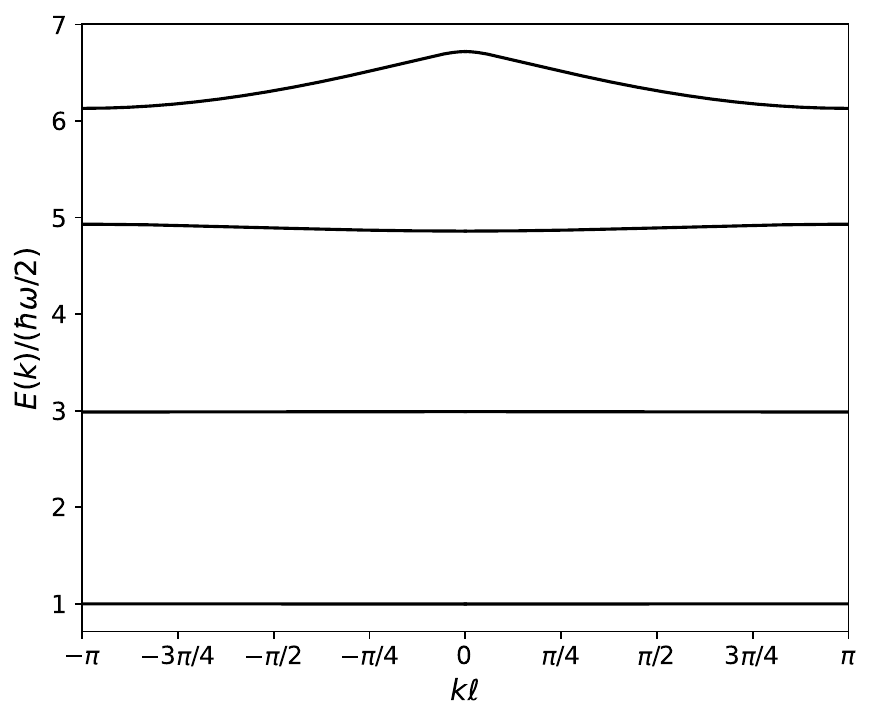}
    \caption{Plot of the lowest four energy bands as a function of wave vector in the first Brillouin zone for the periodic harmonic oscillator potential (see Fig.~\ref{fig:Oscillator Plateau Potential}),
    for \(V_0 / (\hbar\omega / 2) = 6\) and \(w/l = 2/3\). Note how flat the two lowest bands appear on this energy scale; in reality, they have a characteristic cosine-like dispersion, visible when a finer energy scale is used. Also note how close the energy of these two bands is to the lowest two energies of an isolated harmonic oscillator potential, while the third highest band lies entirely below the third level of an isolated harmonic oscillator potential.}
    \label{fig:Bloch Energies}
\end{figure}

As is generically the case, for a deep single well, several bound states will be present. For an infinite array of these wells, each of these levels will spread into bands, and naturally the low-lying bands will be quite narrow, as the tunneling probability will be quite low. In Fig.~\ref{fig:Bloch Energies} we illustrate the energy bands arising in such a case \FM{for the geometry in Fig.~\ref{fig:Oscillator Plateau Potential}.} The lowest two-bands appear flat on this energy scale, but they show the characteristic cosine-like dispersion on a much finer energy scale. These bands are clearly in the tight-binding limit, and a similar case will be discussed in Section~\ref{sec:tight}. The third band shows some dispersion, as well as a reduced energy overall, compared to the isolated harmonic oscillator. This reduction occurs primarily because the harmonic oscillator potential is truncated at an energy just above this band and therefore the electron can tunnel into neighbouring wells.

\begin{figure*}
    \centering
    \includegraphics[width=\textwidth]{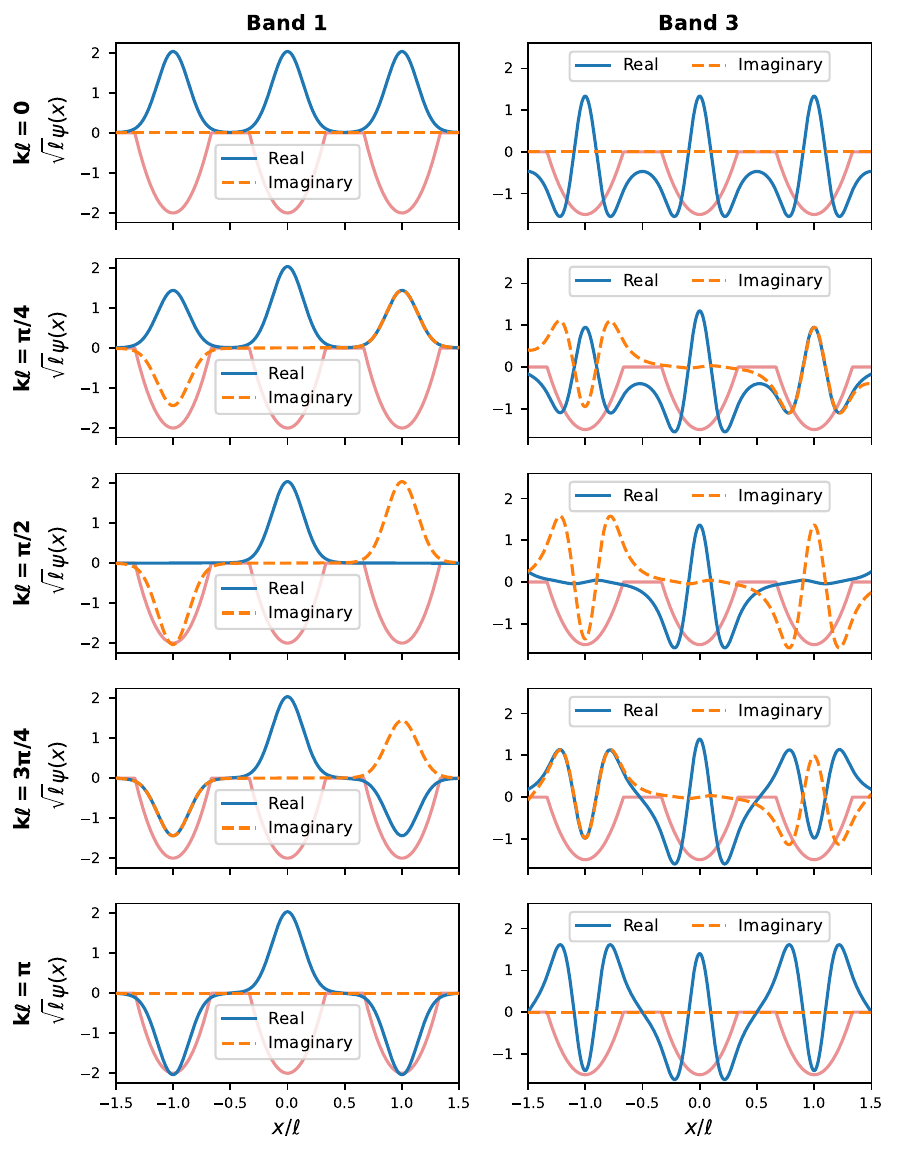}
    \caption{Plot of the real and imaginary parts of the wavefunction as a function of position across three cells for various wave vectors for the first (left panel) and third (right panel) energy bands. These results are shown for the potential illustrated in Fig.~\ref{fig:Oscillator Plateau Potential}, for \(V_0/(\hbar\omega / 2) = 6\) and \(w/l = 2/3\), and this potential is indicated schematically in the figures with a faded red-coloured curve with arbitrary units.}
    \label{fig:Bloch Wavefunctions}
\end{figure*}

Next, we examine the wavefunctions. In Fig.~\ref{fig:Bloch Wavefunctions}, we show the variation with different values of wave vector $k$ in the First Brillouin Zone, for the lowest energy band (left-hand side) and for the third lowest energy band (right-hand side). As the lowest band arises from the ground state for the truncated harmonic oscillator, it will have no structure within the unit cell. However, the variation in the real and imaginary parts will vary from cell to cell according to Eq.~(\ref{eq:bloch_condition}). 

For an energy band with energies closer to the barrier potential, there is considerable structure within the unit cell, as is clear from the right-hand side of Fig.~\ref{fig:Bloch Wavefunctions}. Since this is the third lowest energy band, there will be two nodes within the unit cell, as is most apparent in the $k\ell = 0$ panel. Unlike the wave functions for the lowest energy band (shown on the left), there is now considerable wave function amplitude in the region between potential wells, as the tunneling amplitude is now significantly higher than for the lowest band. 

\section{Tight-Binding Formulation}
\label{sec:tight}
The tight-binding description applies particularly when $\kappa b \gg 1$; that is, we are considering a sufficiently low energy state of a deep well or one separated from neighbouring wells by wide barriers. In one dimension this limit is known to lead to a dispersion relation \cite{Ashcroft_Mermin}
\begin{equation}
\epsilon_{\rm TB}(k) = \epsilon_0 - 2t_1 \cos{(k\ell)} - 2t_2 \cos{(2k\ell)} - \dots
\label{eq:Tight-Binding Dispersion}
\end{equation}
where \(|t_1| \gg |t_2| \gg \dots\). The parameter $t_1$ can be interpreted as a tunnelling amplitude for nearest-neighbour hopping, but in general this interpretation cannot be extended to \(t_2\) and beyond (e.g.\ \(t_2\) cannot be interpreted as an amplitude for next-nearest neighbour hopping) \cite{Kronig-Penney_Tight_Binding}. With the present formulation, one can in fact derive an expression for $t_1$ in terms of the parameters of the periodic truncated harmonic oscillator potential. Equation~(\ref{kp2}) can be rewritten (for $E < V_0$) as
\begin{equation}
f_{\rm even}(\epsilon) f_{\rm odd}(\epsilon) = \eta_1(\epsilon) \cos{(k\ell)} + \eta_2(\epsilon)
\label{eq:Tight Binding Energy}
\end{equation}
where
\begin{equation}
\eta_1(\epsilon) = -2e^{-\kappa b}\left[v_0(1 - \epsilon)m_{53}(\epsilon) - v_0\left(1 - \frac{\epsilon}{3}\right)m_{75}(\epsilon) - 1\right]
\label{eta1}
\end{equation}
and
\begin{equation}
\eta_2(\epsilon) = e^{-2\kappa b}\left[f_{\rm even}(\epsilon) f_{\rm odd}(\epsilon) - 2\right].
\label{eta2}
\end{equation}
Note that Eq.~(\ref{eq:Tight Binding Energy}) is still exact, and that the only $k$-dependence is in the cosine on the right-hand side. As written, both quantities on the right-hand side are very small in the tight-binding limit. Moreover, $\eta_2(\epsilon) \ll \eta_1(\epsilon)$ in this limit. It is clear, for example, that as the potential wells representing the atomic centers are placed further apart, i.e. $b \rightarrow \infty$, then the solution is given by one of the factors on the left-hand side equalling zero, i.e. the even and odd solutions of an isolated well as given in Eqs.~(\ref{feven}) and (\ref{fodd}), respectively. This motivates a perturbative treatment, as was done for the usual Kronig-Penney model in Ref.~\cite{Kronig-Penney_Tight_Binding}.



To see how this comes about, we focus on a bound state with even parity for the isolated well with energy $\epsilon_0$. Therefore $f_{\rm even}(\epsilon_0) = 0$, as given in Eq.~(\ref{feven}). In the present context, $\epsilon_0$ represents the zeroth order solution to Eq.~(\ref{eq:Tight Binding Energy}). For better accuracy, we expand the energy in Eq.~(\ref{eq:Tight Binding Energy}) to first order, $\epsilon_1(k) = \epsilon_0(1 + \rho(k))$, where \mbox{$\rho(k) \ll 1$}. Following Ref.~(\cite{Kronig-Penney_Tight_Binding}), we ignore $\eta_2$
since \mbox{$\eta_2 \ll \eta_1$}.
%
Then one obtains
\begin{equation}
    \rho(k) =
    -\frac{4v_0}{\epsilon_0}\sqrt{1 - \frac{\epsilon_0}{v_0}}\frac{\eta_1(\epsilon_0)}{f_{\rm odd}(\epsilon_0) g(\epsilon_0)} \cos{(k\ell)},
\end{equation}
where
\begin{equation}
g(\epsilon_0) = \frac{2}{\sqrt{1 - \frac{\epsilon_0}{v_0}}} + 4 v_0m_{53}\left(1 + \frac{1}{4}(1-\epsilon_0)(n_{53} - n_{11})\right),
    \label{gdef}
\end{equation}
and we have used 
\begin{equation}
n_{ij} \equiv \frac{1}{M_{ij}} \frac{dM_{ij}}{da}
\label{ndef}
\end{equation}
to denote the logarithmic derivative of the Kummer function with respect to its first parameter $a$ \cite{ancarani_2008}.

This means that the dispersion relation can be written approximately as
\begin{equation}
\epsilon_{\rm TB}(k) = \epsilon_0 - 2t_1 \cos{(k\ell)},
\label{TB_dispersion}
\end{equation}
where
\begin{equation}
t_1 \equiv 2v_0 \sqrt{1 - \frac{\epsilon_0}{v_0}} \frac{\eta_1(\epsilon_0)}{f_{\rm odd}(\epsilon_0) g(\epsilon_0)}
\label{t1_defn}
\end{equation}
is determined completely in terms of the parameters defining the original potential.

%

\begin{figure*}
    \centering
    \includegraphics[width=\textwidth]{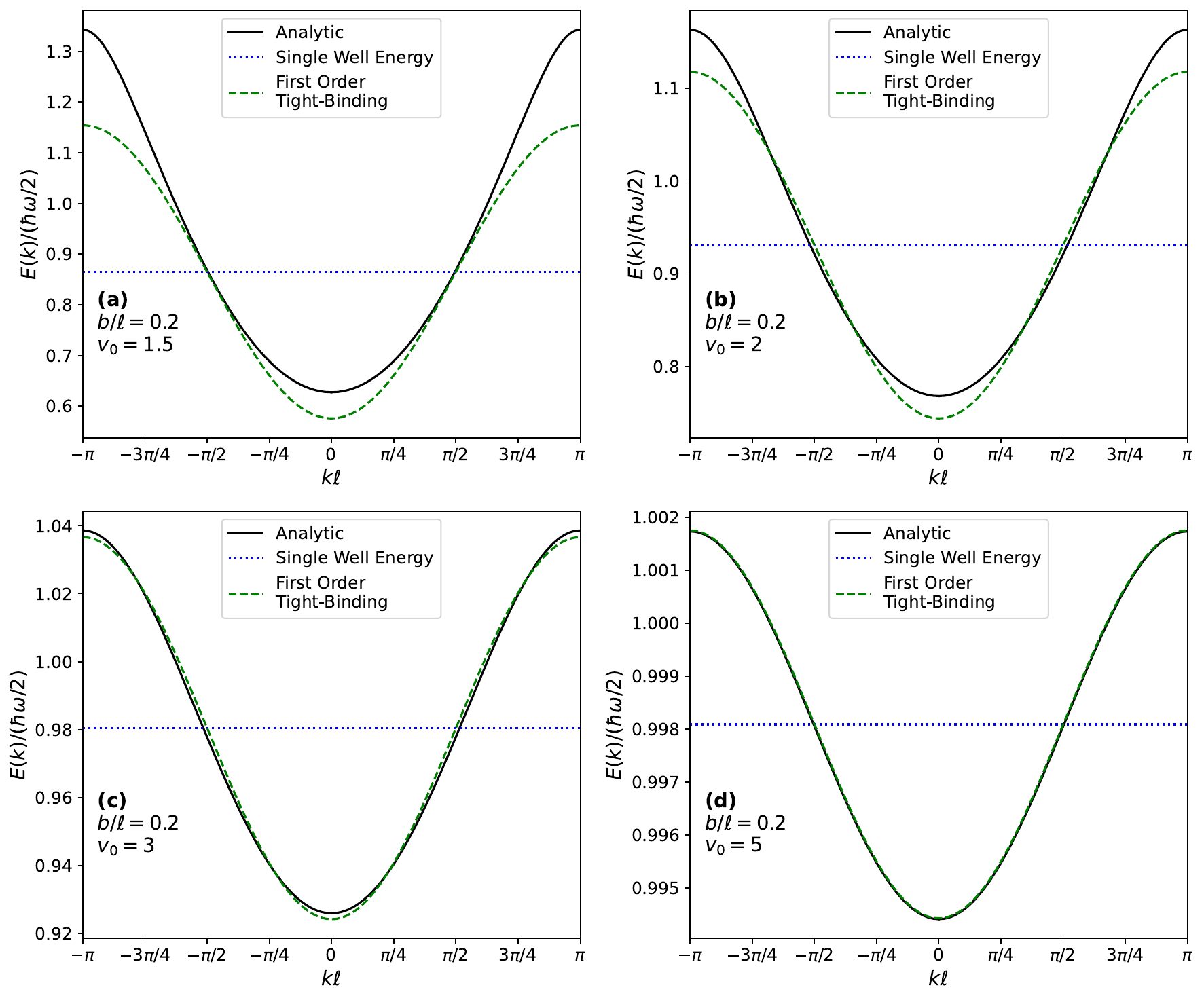}
    \caption{Comparison of the first-order tight-binding dispersion (Equation~\ref{eq:Tight-Binding Dispersion}) with the analytic dispersion for the ground state band. A dashed blue line shows the bound state energy for the single well system; the first-order tight-binding dispersion is automatically symmetric about this energy. All graphs use \(b/l = 1/5\) and left to right, top to bottom the graphs use \(V_0/(\hbar\omega / 2) = 1.5\text{ (a)},\ 2\text{ (b)},\ 3\text{ (c)},\ 5\text{ (d)}\). As the depth of the well is increased the first-order tight-binding dispersion becomes more accurate until at \(V_0 / (\hbar\omega / 2) = 5\) there is visually no difference from the analytic result.}
    \label{fig:First Order Tight Binding Varying v_0}
\end{figure*}

\begin{figure*}
    \centering
    \includegraphics[width=\textwidth]{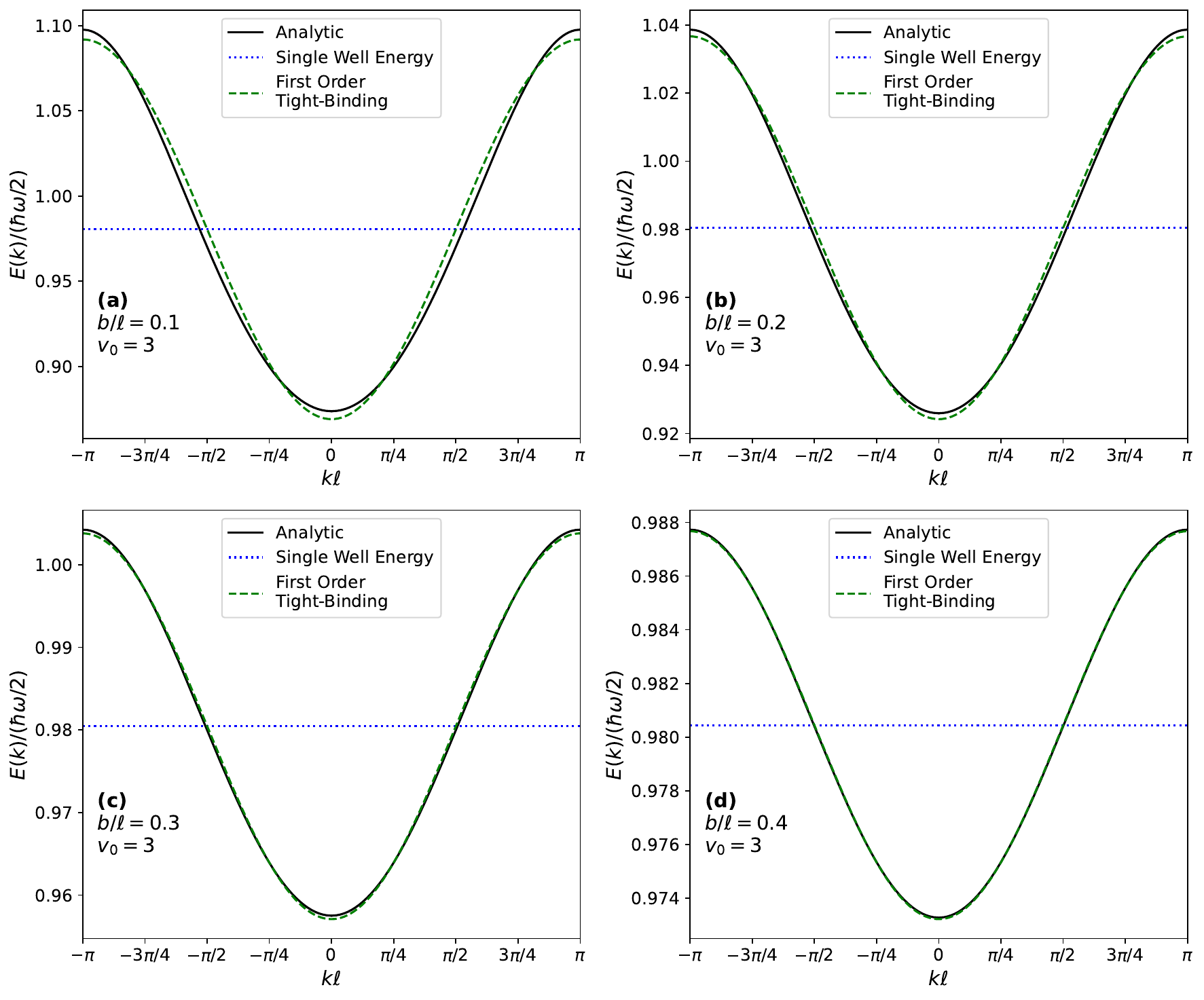}
    \caption{Comparison of the first-order tight-binding dispersion (Equation~\ref{eq:Tight-Binding Dispersion}) with the analytic dispersion for the ground state band. A dashed blue line shows the bound state energy for the single well system; the first-order tight-binding dispersion is automatically symmetric about this energy. All graphs use \(V_0 / (\hbar\omega / 2) = 3\) and left to right, top to bottom the graphs use \(b/l = 0.1\text{ (a)},\ 0.2\text{ (b)},\ 0.3\text{ (c)},\ 0.4\text{ (d)}\). As the width of the barrier is increased the first-order tight-binding dispersion becomes more accurate until at \(b/l = 0.4\) there is visually almost no difference from the analytic result.}
    \label{fig:First Order Tight Binding Varying b_l}
\end{figure*}

Figures~\ref{fig:First Order Tight Binding Varying v_0} and~\ref{fig:First Order Tight Binding Varying b_l} show a comparison for the lowest band energy as a function of wave vector between this result (dashed green line) and the analytic result (solid black line) obtained from solving Eq.~(\ref{kp1}) (or Eq.~(\ref{kp2})
for fixed \(b/\ell\) and different values of \(V_0/(\hbar\omega/2)\) (Figure~\ref{fig:First Order Tight Binding Varying v_0}), and for fixed \(V_0/(\hbar\omega/2)\) and different values of \(b/\ell\) (Figure~\ref{fig:First Order Tight Binding Varying b_l}).

As expected, both figures show that the first-order tight-binding dispersion does not have good agreement with the exact solution when the well is shallow or the barrier is narrow. However, as the height of the well is increased or the width of the barrier is increased, the first-order tight-binding dispersion becomes an excellent approximation of the exact solution. 
In common with the results for the usual Kronig-Penney model (i.e. an array of square wells) in Ref.~\cite{Kronig-Penney_Tight_Binding},
we find that even the first order expansion for the tight-binding model becomes very accurate as the depth of the wells exceeds about $3\times$ the expected isolated well ground state energy. For the present case, this ground state energy is close to $\hbar \omega/2$, whereas in~\cite{Kronig-Penney_Tight_Binding} it is $\hbar^2\pi^2/(2mw^2)$. These results are for $b/\ell = 0.2$, and as 
Fig.~\ref{fig:First Order Tight Binding Varying b_l} clearly shows, the depth of the well can be significantly smaller for large barrier widths, and tight-binding will still be accurate. 

A more meaningful comparison with the results in~\cite{Kronig-Penney_Tight_Binding} is difficult, as the barrier widths quoted for the present case represent the barrier at the top of the potential wells, whereas the effective barrier width for these cases is greater, as can be appreciated by examining the distance between wells not at the top of the wells, but at the level of the state in question, here the lowest band, with energy close to, but below, $\hbar \omega/2$.
The effective width for these levels is then much greater than $b$, the nominal width between wells at their maximum. In contrast, for the usual Kronig-Penney model, this width remains the same, even for the lowest band. This model potential shares this feature in common with the more realistic Coulomb potential that exists in real solids.

One notable difference between the harmonic oscillator and square well potentials is that the exact result in the former case tends to be largely above the single well results. In contrast, for the regular Kronig-Penney model (see, e.g. Fig.~4a in Ref.~\cite{Kronig-Penney_Tight_Binding}), the exact result was below that of a single well. This is the generic result, i.e. coupling to neighbouring wells tends to lower the energy. For the present case, certainly some of the energy levels are lowered below that of the single well, but more levels are raised due to the smaller barriers at higher energy (see, e.g. Fig.~\ref{fig:First Order Tight Binding Varying v_0}(a)).

\section{Conclusion}
\label{conc}
We derived an analytic solution for a Kronig-Penney-like model using a periodic array of harmonic oscillator wells separated by flat barrier regions.
Furthermore, we demonstrated that in the limit of tight-binding, when neighbouring wells are substantially separated, we can obtain an analytic expression for the dispersion relation to first-order (in \(e^{-z}\) where \mbox{\(z \propto b\sqrt{V_0}\)}), that has excellent agreement with the exact solution for modest well heights/barrier widths. Comparing with results obtained for the traditional square-well Kronig-Penney model in~\cite{Kronig-Penney_Tight_Binding} we found that the first-order tight-binding expression works well, in the case of harmonic oscillator wells, for more moderate depths than in the square well case. In addition, in the harmonic oscillator case, the exact solutions are centred around an energy higher than the isolated well energy, which is the opposite of the trend for the square well case studied in~\cite{Kronig-Penney_Tight_Binding}. As was the case in Ref.~\cite{Kronig-Penney_Tight_Binding}, a clear electron-hole asymmetry exists, and for the same reason: states at the top of the band (so-called hole states) are closer to the top of the barrier than states at the bottom of the band (electron states), and so the hole states will have a reduced effective mass, since tunneling is easier for particles in these states. Of course interaction effects, not accounted for here, can (and presumably do) contribute to reverse this trend. 

\begin{acknowledgments}
This work was supported by the Natural Sciences and Engineering Research Council of Canada (NSERC).
\end{acknowledgments}

\appendix

\section{Analytic solutions of the truncated harmonic oscillator potential for various boundary conditions}
\label{sec:Infinite Well Derivation}

We want to solve the potential given by Eq.~(\ref{eq:Oscillator Plateau Potential}), but with a number of auxiliary boundary conditions. These include (i) the flat portions extend to infinity, as written in Eq.~(\ref{eq:Oscillator Plateau Potential}), (ii)  the flat portions stop abruptly at $ = \pm \ell/2$, so that the potential is contained within an infinite square well of width $\ell$, (iii) the same geometry as in (ii), but with periodic boundary conditions, and (iv) the same geometry as in (ii), but with Bloch boundary conditions so that $\psi(x + \ell) = e^{ik\ell} \psi(x)$, as written in Eq.~(\ref{eq:bloch_condition}). This latter boundary condition is the ultimate goal and allows us to solve for the infinite array depicted in Fig.~\ref{fig:Oscillator Plateau Potential}, but by solving for the wave function in one cell only.

To obtain the wave function we must solve the Schr\"{o}dinger equation in the three regions of the potential. In the regions of flat potential we have:

\begin{equation}
    -\frac{\hbar^2}{2m} \dv[2]{\psi}{x} + V_0 \psi = E \psi
\end{equation}

\noindent which leads to hyperbolic (or exponential) solutions (for \(E < V_0\)). Then the general solution in these regions is
\begin{equation}
    \psi(x) = A \exp(\kappa \left(x \pm \frac{w}{2}\right)) + B \exp(-\kappa \left(x \pm \frac{w}{2}\right))
    \label{flat_region_psi}
\end{equation}
where
\begin{equation}
\label{kappa_defn}
    \kappa \equiv\sqrt{\frac{2m(V_0 - E)}{\hbar^2}},
\end{equation}
and one can use either choice indicated in Eq.~(\ref{flat_region_psi}).
For case (i), the requirement that the wave function be square-integrable leads to the choices indicated in the first and third lines of Eq.~\ref{eq:Oscillator Plateau Wavefunction}


For the central region, where the harmonic oscillator potential is present, the Schr\"{o}dinger equation is 
\begin{equation}
\label{sch_harm}
    -\frac{\hbar^2}{2m} \dv[2]{\psi_2}{x} + \frac{1}{2} m \omega^2 x^2 \psi_2 = E \psi_2.
\end{equation}
Following common practice, we adopt dimensionless units, \(z = x / x_0\) and \(\epsilon = E / (\hbar\omega / 2)\), where \(x_0 = \sqrt{\hbar / m\omega}\) and \(\hbar\omega / 2\) are the natural length and energy scales, respectively, for a harmonic oscillator. These substitutions reduce Eq.~(\ref{sch_harm}) to
\begin{equation}
\label{sch_harm_dim}
    \dv[2]{\psi_2}{z} + (\epsilon - z^2) \psi_2 = 0.
\end{equation}
Again, following common practice, we ``peel off'' the asymptotic behaviour, and write \mbox{\(\psi_2(z) \equiv f(z) \exp(-z^2/2)\)}, to get
\begin{equation}
    \dv[2]{f}{z} - 2z \dv{f}{z} + (\epsilon - 1)f = 0.
\end{equation}
Finally, we make the substitution \(u = z^2\), to obtain
\begin{equation}
\label{kummer}
    u \dv[2]{f}{u} + \left(\frac{1}{2} - u\right) \dv{f}{u} - \frac{1 - \epsilon}{4} f = 0.
\end{equation}
This is simply Kummer's equation with \(a = (1 - \epsilon) / 4\) and \(b = 1/2\), and the two linearly independent solutions are~\cite{Handbook_Math_Funcs}
\begin{equation*}
    \kummer{\frac{1 - \epsilon}{4}}{\frac{1}{2}}{u}\ \text{and}\ \sqrt{u}\kummer{\frac{3 - \epsilon}{4}}{\frac{3}{2}}{u},
\end{equation*}
where
\begin{equation}
    M(a, b, u) = \sum_{n=0}^{\infty} \frac{(a)_n}{(b)_n} \frac{u^n}{n!}
\end{equation}
is the Kummer function, and
\begin{equation}
    \label{pochhammer}
    (a)_n \equiv a(a+1)(a+2)\cdots (a+n-1), \ \ \ (a)_0 \equiv 1,
\end{equation}
is Pochhammer's symbol~\cite{Handbook_Math_Funcs}.

\begin{widetext}
These observations result in the most general solution to Eq.~\ref{sch_harm},
\begin{equation}
\label{sch_gen}
    \psi_2(x) =\left[
        C \kummer{\frac{1 - \epsilon}{4}}{\frac{1}{2}}{\left(\frac{x}{x_0}\right)^2}
        + D \frac{x}{x_0} \kummer{\frac{3 - \epsilon}{4}}{\frac{3}{2}}{\left(\frac{x}{x_0}\right)^2}\right]
    e^{-(x / x_0)^2 / 2}
\end{equation}
\end{widetext}
with $C$ and $D$ arbitrary coefficients. 
This wave function is applicable in the harmonic oscillator region, $-w/2~<~x~<~w/2$. Note that the Kummer functions reduce to Hermite polynomials when the first argument is a negative integer~\cite{Handbook_Math_Funcs}, consistent with the dimensionless energies being proportional to an odd integer, as occurs for the case of an infinite harmonic oscillator. Instead, here the energy is determined by matching the wave function and its derivative at $x = \pm w/2$, and the more general form of the Kummer function is therefore required.

Combining these solutions results in the wave function written in Eq.~\ref{eq:Oscillator Plateau Wavefunction}.
Consideration of the symmetry leads to considerable simplifications. 
We expect both even (\(F=A\), \(D=0\)) and odd (\(F=-A\), \(C=0\)) solutions. For these two cases we match boundary conditions at \(x = w/2\), which immediately results in the equations
\begin{equation}
f_{\rm even}(\epsilon) \equiv 1 - \frac{1 - (1-\epsilon)m_{53}(\epsilon)}{\sqrt{1 - {\epsilon}/{v_0}}} = 0 \ \ \ \ \ \ \ \ \ \ \ \ \ \ \ \ \ {\rm (even) \ \ \ }
\label{feven_app}
\end{equation}
and
\begin{equation}
f_{\rm odd}(\epsilon) \equiv 1 - v_0 + v_0\sqrt{1 - \frac{\epsilon}{v_0}} + v_0(1 - \frac{\epsilon}{3})m_{75}(\epsilon) = 0 \ \ \ {\rm (odd),}
\label{fodd_app}
\end{equation}
as written in Eqs.~(\ref{feven}) and (\ref{fodd}).
Note that we have made use of the identity \cite{Handbook_Math_Funcs} \begin{equation}
    \label{kummer_deriv}
\frac{dM(a,b,z)}{dz} = \frac{a}{b} M(a+1,b+1,z),
\end{equation}
and we remind the reader that
\begin{equation}
    \label{remind} 
    m_{ij}(\epsilon) \equiv \frac{M_{ij}(\epsilon)}{M_{i-4,j-2}(\epsilon)} \ \ {\rm and} \ \  M_{ij}(\epsilon) \equiv M\left(\frac{i-\epsilon}{4},\frac{j}{2},v_0\right).
\end{equation}

For the next case considered, a truncated harmonic oscillator potential embedded in an infinite potential well of width $\ell$, Eq.~(\ref{eq:Oscillator Plateau Potential}) is modified to the form
\begin{equation}\label{eq:Oscillator Plateau Potential_inf}
    V_\mathrm{HO}(x) = V_0\begin{cases}
        1 & \text{if } \  \  -\ell/2 < x \leq -w/2 \\
        (\frac{x}{w/2})^2 & \text{if }\ -w/2 \leq x \leq w/2 \ \\
        1 & \text{if }\ \ \ \ \ \ w/2 \leq x < \ell/2 \\
        \infty & {\rm otherwise.}
    \end{cases}
\end{equation}
Everything proceeds as before, except the left-most region now has a wave function solution of
\begin{equation}
    \label{psi_left}
    \psi_{\rm left}(x) =
        A \sinh(\kappa \left(x + \frac{\ell}{2}\right)),
\end{equation}
and the right-most region has a wave function solution of
\begin{equation}
    \label{psi_left}
    \psi_{\rm right}(x) =
        \pm A \sinh\left(\kappa \left(\frac{\ell}{2} - x\right)\right),
\end{equation}
depending on whether we adopt the even (+) or odd (-) solution, respectively. These are matched with the corresponding even and odd solution, respectively, in the central region, to obtain equations determining the allowed energies for even or odd states,
\begin{widetext}
\begin{equation}
    \label{feven_inf_well}
            1 
-\frac{\tanh\left(\frac{\kappa b}{2}\right)[1 - (1 - \epsilon)m_{53}(\epsilon)]}{\sqrt{1 - \epsilon/v_0}} = 0, \quad \quad  \quad \quad \quad \quad \quad \quad \quad \quad \quad \quad \quad \quad \quad  {\rm (even \ bound \ states, \ Dirichlet \ BCs)}
    \end{equation}
and
    \begin{equation}
    \label{fodd_inf_well}
            (1 - v_0)\tanh\left(\frac{\kappa b}{2}\right)
            + v_0\sqrt{1 - \frac{\epsilon}{v_0}} + v_0\left(1 - \frac{\epsilon}{3}\right)
    \tanh\left(\frac{\kappa b}{2}\right)
            m_{75}(\epsilon) = 0 \ \quad  {\rm (odd \ bound \ states, \ Dirichlet \ BCs)},
\end{equation}
\end{widetext}
where 
\begin{equation}
    \label{kappa_b_defn}
    \tanh\left(\frac{\kappa b}{2}\right) \equiv \tanh\left(\frac{b}{2x_0}\sqrt{v_0 - \epsilon}\right).
\end{equation}
Solutions to these equations are readily obtained; for typical parameters considered in this study, the differences arising from the infinite square well embedding potential are minor.

Note that a similar procedure 
can be executed for states with energy levels $E > V_0$. Then, one defines $Q\equiv \sqrt{\frac{2m}{\hbar^2} (E - V_0)}$ and obtains
\begin{widetext}
\begin{equation}
    \label{feven_inf_well_scat}
            1 
-\frac{\tan\left(\frac{Q b}{2}\right)[1 - (1 - \epsilon)m_{53}(\epsilon)]}{\sqrt{\epsilon/v_0} - 1} = 0 \ \quad \quad \quad \quad \quad \quad \quad \quad \quad \quad \quad \quad \quad \quad \quad \quad \quad \quad  {\rm (even \ states \ Dirichlet \ \ } E > V_0)
    \end{equation}
and
    \begin{equation}
    \label{fodd_inf_well_scat}
            (1 - v_0)\tan\left(\frac{Q b}{2}\right)
            + v_0\sqrt{ \frac{\epsilon}{v_0} - 1} + v_0\left(1 - \frac{\epsilon}{3}\right)
    \tan\left(\frac{Q b}{2}\right)
            m_{75}(\epsilon) = 0 \ \quad \quad \quad  {\rm (odd \ states,\ Dirichlet \ \ } E > V_0),
\end{equation}
where 
\begin{equation}
    \label{kappa_b_defn}
    \tan\left(\frac{Q b}{2}\right) \equiv \tan\left(\frac{b}{2x_0}\sqrt{ \epsilon - v_0}\right).
\end{equation}

In the case of periodic boundary conditions,
for $E<V_0$, we can first write the most general wave function as
\begin{equation}
    \psi(x) = \begin{cases}
        A \sinh(\kappa \left(x + \frac{\ell}{2}\right)) + B \cosh(\kappa \left(x + \frac{\ell}{2}\right)) & \text{if } -\ell/2 \leq x \leq -w/2 \\
        \left[C \kummer{\frac{1 - \epsilon}{4}}{\frac{1}{2}}{[x/x_0]^2} + D \frac{x}{x_0} \kummer{\frac{3 - \epsilon}{4}}{\frac{3}{2}}{[x/x_0]^2}\right] e^{-[x/x_0]^2/2} & \text{if } -w/2 \leq x \leq w/2 \\
        F \sinh(\kappa \left(\frac{\ell}{2} - x\right)) + G \cosh(\kappa \left(\frac{\ell}{2} - x\right)) & \text{if } \ \ \ \ w/2 \leq x \leq \ell/2, \\
    \end{cases}
\end{equation}
\end{widetext}
and then apply the periodic boundary conditions, \(\psi(-l/2) = \psi(l/2)\) and \(\psi'(-l/2) = \psi'(l/2)\).  Even with periodic boundary conditions, we find that using the matching conditions at $x = \pm w/2$ decouples the four remaining equations into even ($A = F = D = 0$) and odd ($B = G = C = 0$) solutions. The resulting equations are
\begin{widetext}
\begin{equation}
    \label{feven_periodic_well}
            1 
-\frac{\coth\left(\frac{\kappa b}{2}\right)[1 - (1 - \epsilon)m_{53}(\epsilon)]}{\sqrt{1 - \epsilon/v_0}} = 0 \ \quad \quad \quad \quad \quad \quad \quad \quad \quad \quad \quad \quad \quad \quad \quad \quad  {\rm (even \ bound \ states, \ periodic \ BCs)}
    \end{equation}
and
    \begin{equation}
    \label{fodd_periodic_well}
            (1 - v_0)\tanh\left(\frac{\kappa b}{2}\right)
            + v_0\sqrt{1 - \frac{\epsilon}{v_0}} + v_0\left(1 - \frac{\epsilon}{3}\right)
    \tanh\left(\frac{\kappa b}{2}\right)
            m_{75}(\epsilon) = 0 \ \quad \quad \quad  {\rm (odd \ bound \ states, \ periodic \ BCs)}.
\end{equation}
\end{widetext}
Note that the equation for odd bound state energies is identical to that for Dirichlet boundary conditions, Eq.~(\ref{fodd_inf_well}), while the even bound state equation has a $coth$ function in place of a $tanh$ function, but is otherwise the same as Eq.~(\ref{feven_inf_well}). A similar difference occurs for the usual square well model, and reflects the fact that with Dirichlet boundary conditions, both even and odd bound states have higher energy compared to the free case, due to the quantum pressure exerted by the walls, whereas with periodic boundary conditions, the impact of the periodicity is identical to that of infinite walls for odd states, whereas for even states, periodic boundary conditions provide more freedom to the bound state wave function to evanesce, and therefore the energy decreases in comparison to the case with infinite walls. 

The scattering states ($E > 0$) work as before, and with periodic boundary conditions we obtain the following equations
\begin{widetext}
\begin{equation}
    \label{feven_periodic_scat}
            1 
+\frac{\cot\left(\frac{Q b}{2}\right)[1 - (1 - \epsilon)m_{53}(\epsilon)]}{\sqrt{\epsilon/v_0 - 1}} = 0 \ \quad \quad \quad \quad \quad \quad \quad \quad \quad \quad \quad \quad \quad \quad \quad \quad \quad {\rm (even \ states \ periodic \ \ } E > V_0)
    \end{equation}
and
    \begin{equation}
    \label{fodd_periodic_scat}
            (1 - v_0)\tan\left(\frac{Q b}{2}\right)
            + v_0\sqrt{ \frac{\epsilon}{v_0} - 1} + v_0\left(1 - \frac{\epsilon}{3}\right)
    \tan\left(\frac{Q b}{2}\right)
            m_{75}(\epsilon) = 0 \ \quad \quad \quad  {\rm (odd \ states,\ periodic \ \ } E > V_0),
\end{equation}
\end{widetext}
The equation for the even solutions is altered slightly, whereas the equation for the odd solutions is identical to Eq.~(\ref{fodd_inf_well_scat}) for the Dirichlet boundary conditions.

For the Bloch boundary condition,
we start with the general wave function, Eq.~(\ref{eq:Oscillator Plateau Full Wavefunction}), and apply the Bloch conditions, \(\psi(\ell/2) = e^{ik\ell} \psi(-\ell/2)\) and \(\psi'(\ell/2) = e^{ik\ell} \psi'(-\ell/2)\), where \(k\) is the Bloch wavenumber~\cite{Bloch}. These conditions require \(F = -e^{ik\ell} A\) and \(G = e^{ik\ell} B\). The we match boundary conditions at \(x = \pm w/2\) to get four equations. Taking judicious linear combinations of these equations results in the following four equations,
\begin{widetext}
\begin{align}
\label{four_equations}
B\cosh{\kappa b \over 2}&= \ \ CM_{11}\cos{k\ell \over 2} &-iD\sqrt{v_0} M_{33}\sin{k\ell \over 2}\\
A\sinh{\kappa b \over 2}&= iCM_{11} \sin{k\ell \over 2} &- D\sqrt{v_0} M_{33} \cos{k\ell \over 2} \\
\kappa x_0 \ A\cosh{\kappa b \over 2} &= iC\sqrt{v_0} \left(M_{11} - (1-\epsilon)M_{53}\right)\sin{k\ell \over 2} &+D\left[M_{33}(1-v_0) + v_0 (1 - {\epsilon \over 3})M_{75}\right] \cos{k\ell \over 2}\\
\kappa x_0 \ B\sinh{\kappa b \over 2} &= \ C\sqrt{v_0} \left(M_{11} - (1-\epsilon)M_{53}\right)\cos{k\ell \over 2} &+iD\left[M_{33}(1-v_0) + v_0 (1 - {\epsilon \over 3})M_{75}\right] \sin{k\ell \over 2}.
\end{align}
\end{widetext}
In constructing these equations we have redefined some of the coefficients, for example $Ae^{ik\ell/2} \rightarrow A$, and similarly for $B$, and $C \exp{(-{v_0/2})} \rightarrow C$, and likewise for $D$. Demanding consistency between the two equations for $A$ and similarly for $B$ results in two linear homogeneous equations, and therefore requires the evaluation of a $2\times 2$ matrix determinant. The result is given in Eq.~(\ref{kp1}). 

For $E> V_0$, a similar equation results, essentially with $\kappa b \rightarrow iQb$.

In all cases, energies are given by the zeros of these non-linear equations. In practice the equations are solved by stepping through values of $\epsilon$ until a zero-crossing is observed, and then we refine the location of the zero to a desired accuracy. Note that when solving the equations we found it easiest to deal with the equation with no $M_{ij}$ in the denominators (i.e. by not using the $m_{ij}$ as defined right after Eq.~{\ref{fodd}). This avoids spurious zeros arising from the poles of the Kummer functions. As expected, the numerically obtained energies are in excellent agreement with the analytically obtained energies; the difference of the numerical results with the analytical ones is typically of the order of \(\sim10^{-12}\). 

\section{Numerical solutions}
\label{sec:Numerical Solutions}

By embedding the potential in an infinite square well it becomes relatively straightforward to derive numerical results for the bound state energies and wave functions. The methodology is fully described in Ref.~\cite{Kronig-Penney_Matrix_Mechanics}, but the key idea is to expand the wave function in the basis of the infinite square well eigenstates. The specific form of these basis states depends on the boundary conditions used.

For Dirichlet boundary conditions the expansion is given by
\begin{equation}
    \ket{\psi} = \sum_{n=1}^{\infty} c_n \ket{\psi_n^{(0)}},
\end{equation}
where the basis states are
\begin{equation}
    \psi_n^{(0)}(x) = \begin{cases}
        \sqrt{\frac{2}{\ell}} \sin(\frac{n\pi}{\ell} \left(x + \frac{\ell}{2}\right)) & \text{if } -\ell/2 \leq x \leq \ell/2 \\
        0 & \text{otherwise}.
    \end{cases}
\end{equation}
With no potential, these eigenstates have eigenvalues
\begin{equation}
    E_n^{(0)} = \frac{n^2 \pi^2 \hbar^2}{2m\ell^2} \equiv n^2 E_1^{(0)}.
\end{equation}

On the other hand, for periodic boundary conditions the expansion goes over positive and negative integers,
\begin{equation}
    \ket{\psi} = \sum_{n=-\infty}^{\infty} c_n \ket{\psi_n^{(0)}}
\end{equation}
where the basis states are
\begin{equation}
    \psi_n^{(0)}(x) = \begin{cases}
        \frac{1}{\sqrt{\ell}} \exp(i\frac{2n\pi}{\ell} x) & \text{if } -\ell/2 \leq x \leq \ell/2 \\
        0 & \text{otherwise}
    \end{cases}
\end{equation}
with eigenvalues
\begin{equation}
    E_n^{(0)} = 4n^2 E_1^{(0)} = 4 \frac{n^2 \pi^2 \hbar^2}{2m\ell^2}.
\end{equation}
Finally, for Bloch boundary conditions the expansion is the same as for the periodic boundary condition case except that
\begin{equation*}
    \frac{2n\pi}{\ell} \rightarrow \frac{2n\pi}{\ell} + k
\end{equation*}
in the basis states and
\begin{equation*}
    4n^2 \rightarrow \left(2n + \frac{k\ell}{\pi}\right)^2
\end{equation*}
in the eigenvalues where $k$ is the Bloch wave vector.

Using the appropriate expansion we then reduce the problem of finding the allowed energies and associated wavefunctions to a matrix diagonalization problem

\begin{equation}
    \sum_{m} H_{nm} c_m = Ec_n
\end{equation}

\noindent where

\begin{align}
    H_{nm} &= \matrixel{\psi_n}{\hat{H}}{\psi_m} \\
           &= E_n^{(0)}\delta_{nm} + \int_{-\ell/2}^{\ell/2} \psi_n(x) V_\mathrm{HO}(x) \psi_m(x) \dd{x} \nonumber
\end{align}

\noindent and $V_\mathrm{HO}(x)$ is as defined by Eq.~\ref{eq:Oscillator Plateau Potential}. Note that going forward we will work with dimensionless energy in units of $\hbar\omega / 2$ where $h_{nm} \equiv H_{nm} / (\hbar\omega / 2)$, $\epsilon \equiv E / (\hbar\omega / 2)$, and $e_n^{(0)} \equiv E_n^{(0)} / (\hbar\omega / 2)$. Furthermore, to handle the negative values of $n$ that occur for periodic and Bloch boundary conditions we order the rows and columns of the Hamiltonian matrix as \(n' = \{0,-1,+1,-2,+2,\dots\}\) but map these to \(n = \{1,2,3,4,5,\dots\}\) when plotting or comparing with other results.

Proceeding with the calculation of the matrix for Dirichlet boundary conditions gives

\begin{equation}\label{eq:Inf Well Matrix Elements}
\begin{split}
    h_{nm} &=
        \delta_{nm} \left[
            n^2 e_1^{(0)} + v_0 \left(
                1 - \frac{2}{3}\frac{w}{\ell}
                - \frac{2}{n^2\pi^2} \frac{\ell^2}{w^2} f_n
            \right)
        \right] \\
        & \qquad + (1 - \delta_{nm}) 8 v_0 \frac{\ell^2}{w^2} \left(\frac{(-1)^{n+m} + 1}{2}\right) g_{nm}
\end{split}
\end{equation}

\noindent where

\begin{align}
    f_n &= \frac{w}{\ell} \cos(n\pi \frac{b}{\ell}) + \frac{1}{n\pi} \sin(n\pi \frac{b}{\ell}) \ \ \ \ \ \  {\rm and}\\
\begin{split}
    g_{nm} &= 
            \frac{2}{\pi^3} \left( \frac{\sin((n-m)\frac{\pi}{2} \frac{b}{\ell})}{(n-m)^3} - \frac{\sin((n+m)\frac{\pi}{2} \frac{b}{\ell})}{(n+m)^3} \right) \\
            & + \frac{w/\ell}{\pi^2} \left( \frac{\cos((n-m)\frac{\pi}{2} \frac{b}{\ell})}{(n-m)^2} - \frac{\cos((n+m)\frac{\pi}{2} \frac{b}{\ell})}{(n+m)^2}  \right).
\end{split}
\end{align}
For periodic boundary conditions, we obtain
\begin{equation}\label{eq:Periodic BC Matrix Elements}
\begin{split}
    h_{nm} &=
        \delta_{nm} \left[
            4n^2 e_1^{(0)} + v_0 \left(
                1 - \frac{2}{3}\frac{w}{\ell}
            \right)
        \right] \\
        & \qquad + (1 - \delta_{nm}) \frac{2 v_0}{\pi^2 (n - m)^2} \frac{\ell}{w} p_{nm}
\end{split}
\end{equation}
where
\begin{equation}
    p_{nm} =
        \cos((n-m)\pi\frac{w}{\ell}) - {\rm sinc}((n-m)\pi\frac{w}{\ell}), 
\end{equation}
and ${\rm sinc}(x) \equiv \sin{x}/x$.
\noindent Note that the Bloch case is easily determined from the periodic case, as previously mentioned.

\begin{figure}
    \centering
    \includegraphics[width=\columnwidth]{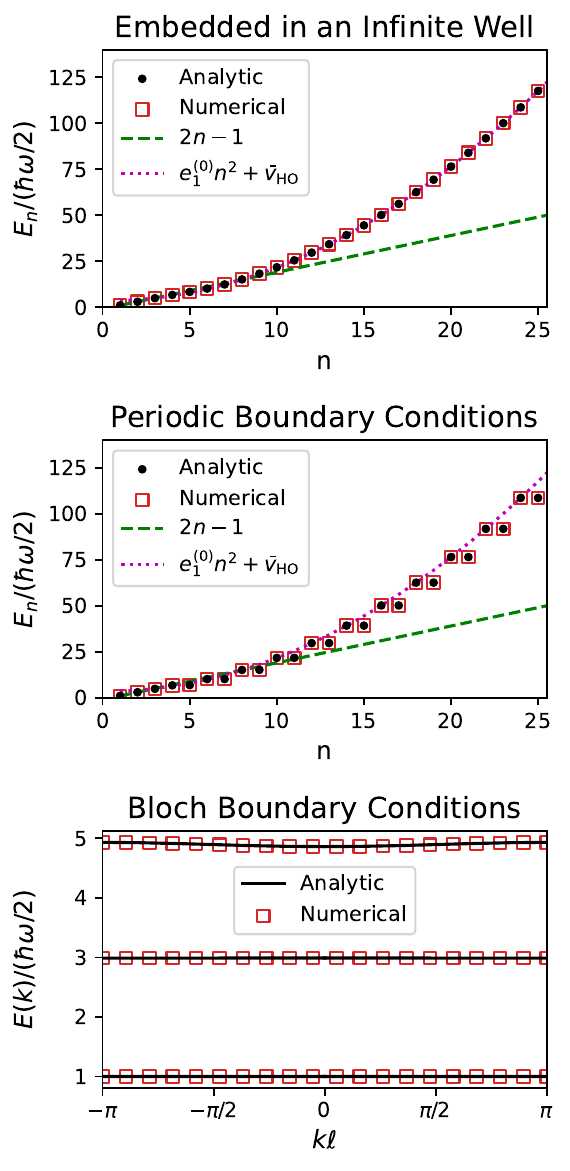}
    \caption{Plot of the first few energy levels of the harmonic oscillator well potential (see Fig.~\ref{fig:Oscillator Plateau Potential}) with \(V_0 / (\hbar\omega / 2) = 6\) and \(w/\ell = 2/3\). From top to bottom the figures show the energy levels when embedded in an infinite square well, with periodic boundary conditions, and with Bloch boundary conditions. For the Bloch boundary condition case the first three energy bands as a function of the wave vector in the first Brillouin zone are shown. In all cases both numerical and analytic energies are shown. At low energies the allowed energies follow those of a harmonic oscillator (dashed green line/odd integers), and at large energies the allowed energies follow those of an infinite square well offset by the average of the embedded potential (dotted pink curve). Additionally, for the Bloch boundary condition case note how flat the two lowest bands appear on this energy scale; in reality, they have a characteristic cosine-like dispersion on a much lower energy scale (see Fig.~\ref{fig:Numerical Bloch Bands}).}
    \label{fig:Numerical Eigenvalues}
\end{figure}

\begin{figure}
    \centering
    \includegraphics[width=\columnwidth]{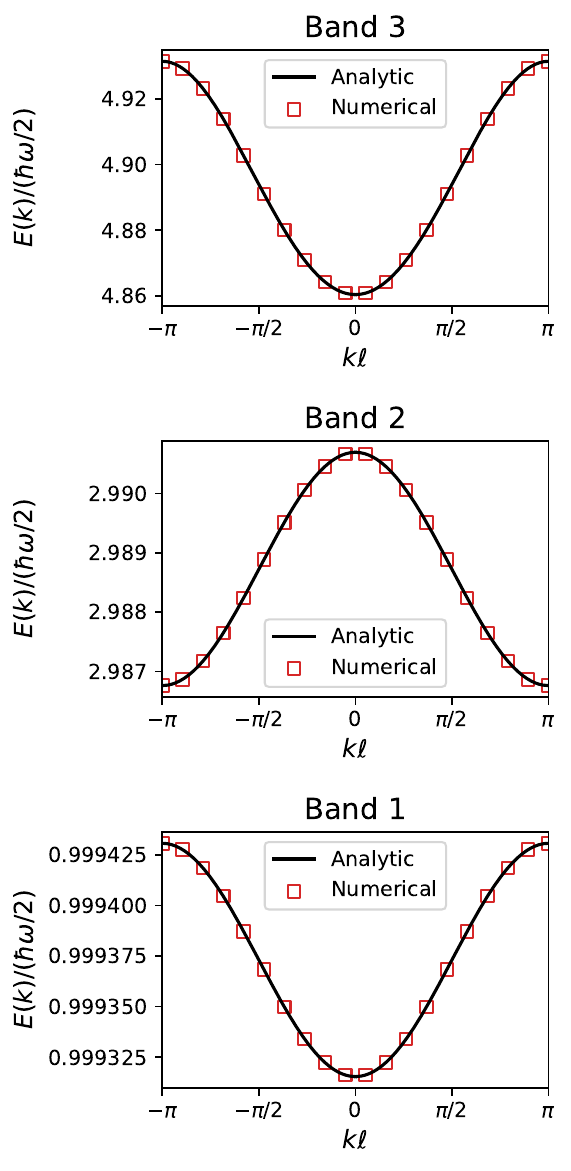}
    \caption{Expanded plot of the first three energy bands in the first Brillouin zone from Fig.~\ref{fig:Numerical Eigenvalues} for the harmonic oscillator well potential (see Fig.~\ref{fig:Oscillator Plateau Potential}) with \(V_0 / (\hbar\omega / 2) = 6\), \(w/\ell = 2/3\), and Bloch boundary conditions. Both numerically (red squares) and analytically (black line) obtained energies are shown. All bands show similar dispersion, albeit on very different energy scales, despite all appearing flat in Fig.~\ref{fig:Numerical Eigenvalues}.}
    \label{fig:Numerical Bloch Bands}
\end{figure}

In Fig.~\ref{fig:Numerical Eigenvalues} we show the first few allowed energies obtained by diagonalizing the Hamiltonian matrix for each of the three cases, (i) embedded in an infinite square well, (ii) with periodic boundary conditions, and (iii) with Bloch boundary conditions (see also Fig.~\ref{fig:Numerical Bloch Bands}). Note that we used a truncated matrix of size $400 \times 400$ ($n=1$ to $n=400$) for the infinite well case and $401 \times 401$ ($n=-200$ to $n=200$) for the periodic and Bloch cases. These numbers can be easily increased by an order of magnitude or more, so that the acquired accuracy can become machine precision.

For comparison, we also plot the energy levels for a pure harmonic oscillator, \(E_n = \hbar\omega(n - 1/2)\), and the energy levels for a pure infinite square well offset by the average of the potential, \(E_n = n^2 E_1^{(0)} + \bar{V}_\mathrm{HO}\). This offset,

\begin{equation}\label{eq:Potential Average}
    \bar{v}_0 = \frac{\bar{V}_0}{\hbar\omega / 2} = v_0\left(1 - \frac{2}{3}\frac{w}{\ell}\right)
\end{equation}

\noindent is deduced by inspection of Eq.~\ref{eq:Inf Well Matrix Elements} or~\ref{eq:Periodic BC Matrix Elements} for large \(n\), and physically corresponds to the fact that states at very high energy see a small potential that partially fills the bottom of the infinite square well.

At low energies we observe that the energy levels closely follow those of a harmonic oscillator as expected due to the exponential decay of the wavefunction away from \(x = \pm w/2\). Furthermore, at large energies when the embedded potential is effectively a constant offset to the floor of the infinite well we observe that the energy levels closely follow those of an infinite square well but with the aforementioned offset (see Eq.~\ref{eq:Potential Average}). Note that in the case of periodic boundary conditions the energies become doubly degenerate at large \(n\) because there are two solutions for each energy, a right-moving plane wave and a left-moving plane wave. 

It is also important to note that while the two lowest bands shown in the Bloch boundary condition case appear flat on this energy scale, they show the characteristic cosine-like dispersion on a much finer energy scale, see Fig.~\ref{fig:Numerical Bloch Bands}. These bands are clearly in the tight-binding limit, and a similar case is discussed in Section~\ref{sec:tight}.


We can compare the wave functions corresponding to the different boundary conditions. In Fig.~\ref{fig:Numerical Wavefunctions} we show the ground state wave functions for (i) the case of being embedded in an infinite square well and (ii) for the case of periodic boundary conditions. For comparison, the wave function for the ground state harmonic oscillator
\begin{equation}
    \psi_\mathrm{HO}(x) =
    \left(\frac{1}{\pi x_0^2}\right)^{1/4}
    e^{-[x/x_0]^2 / 2}
\end{equation}
is also shown. As expected for these parameters, the ground state wave functions for both cases agree very well with the harmonic oscillator ground state wave function because the ground state energy is well below the barrier potential. The maximum difference in Fig.~\ref{fig:Numerical Wavefunctions} is \(\sim 10^{-2}\), which can be arbitrarily reduced by increasing the barrier height.

\begin{figure}
    \centering
    \includegraphics[width=\columnwidth]{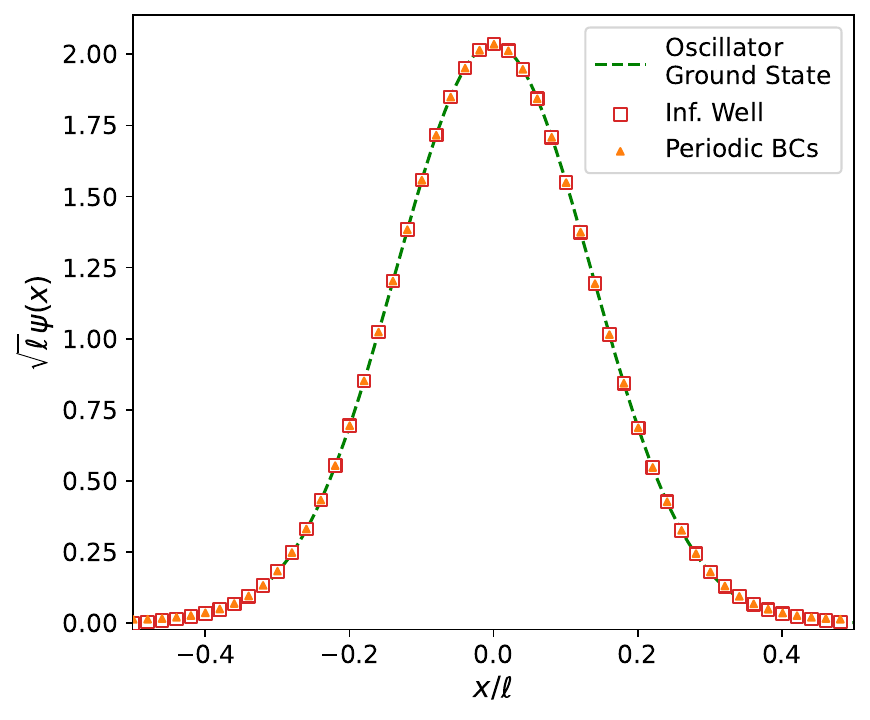}
    \caption{Plot of the numerical ground state wave function for the harmonic oscillator well potential (see Fig.~\ref{fig:Oscillator Plateau Potential}) with Dirichlet and periodic boundary conditions for \(V_0 / (\hbar\omega / 2) = 6\) and \(w/\ell = 2/3\). For comparison the wave function of a ground state harmonic oscillator is also shown. As expected the ground state wave functions of the well potential agree with the harmonic oscillator ground state wave function because the ground state energy is well below the barrier.}
    \label{fig:Numerical Wavefunctions}
\end{figure}

\bibliography{paper}

\end{document}